\documentclass[letterpaper]{article} 
\usepackage[]{aaai24}  
\usepackage{times}  
\usepackage{helvet}  
\usepackage{courier}  
\usepackage[hyphens]{url}  
\usepackage{graphicx} 
\usepackage{natbib}  
\usepackage{caption} 
\usepackage{algorithm}
\usepackage{algorithmic}
\usepackage{xcolor}
\usepackage{dblfloatfix}
\usepackage{newfloat}
\usepackage{listings}
\DeclareCaptionStyle{ruled}{labelfont=normalfont,labelsep=colon,strut=off} 
\floatstyle{ruled}
\newfloat{listing}{tb}{lst}{}
\floatname{listing}{Listing}
\title{Automatic Screening for Children with Speech Disorder using Automatic Speech Recognition: Opportunities and Challenges}
\author{
    Dancheng Liu\textsuperscript{\rm 1}, 
    Jason Yang\workdone\textsuperscript{\rm 1},
    Ishan Albrecht-Buehler\equalcontrib\workdone\textsuperscript{\rm 1,\rm 2},
    Helen Qin\equalcontrib\workdone\textsuperscript{\rm 1,\rm 3},
    Sophie Li\workdone\textsuperscript{\rm 1,\rm 4}\\
    Yuting Hu\textsuperscript{\rm 1},
    Amir Nassereldine\textsuperscript{\rm 1},
    Jinjun Xiong\textsuperscript{\rm 1}
}
\affiliations{
    \textsuperscript{\rm 1}University at Buffalo\\
    \textsuperscript{\rm 2}Nichols School\\
    \textsuperscript{\rm 3}Thomas Jefferson High School for Science \& Technology\\
    \textsuperscript{\rm 4}Langley High School\\
    \textsuperscript{\rm 1}\{dliu37, jyang86, yhu54, amirnass, jinjun\}@buffalo.edu, 	\textsuperscript{\rm 2}ishanalbrecht@outlook.com \\ \textsuperscript{\rm 3}helen.ty.qin@gmail.com,	 \textsuperscript{\rm 4}sophie.zhiran.li@gmail.com
    
}

\usepackage{bibentry}

\begin{document}

\maketitle

\begin{abstract}
Speech is a fundamental aspect of human life, crucial not only for communication but also for cognitive, social, and academic development. 
Children with speech disorders (SD) face significant challenges that, if unaddressed, can result in lasting negative impacts. 
Traditionally, speech and language assessments (SLA) have been conducted by skilled speech-language pathologists (SLPs), but there is a growing need for efficient and scalable SLA methods powered by artificial intelligence. This position paper presents a survey of existing techniques suitable for automating SLA pipelines, with an emphasis on adapting automatic speech recognition (ASR) models for children's speech, an overview of current SLAs and their automated counterparts to demonstrate the feasibility of  AI-enhanced SLA pipelines, and a discussion of practical considerations, including accessibility and privacy concerns, associated with the deployment of AI-powered SLAs. 
\end{abstract}
\begin{figure*}[!btp]
  \centering
  \includegraphics[clip,trim={0cm 5cm 0cm 0.9cm},width=0.9\textwidth]{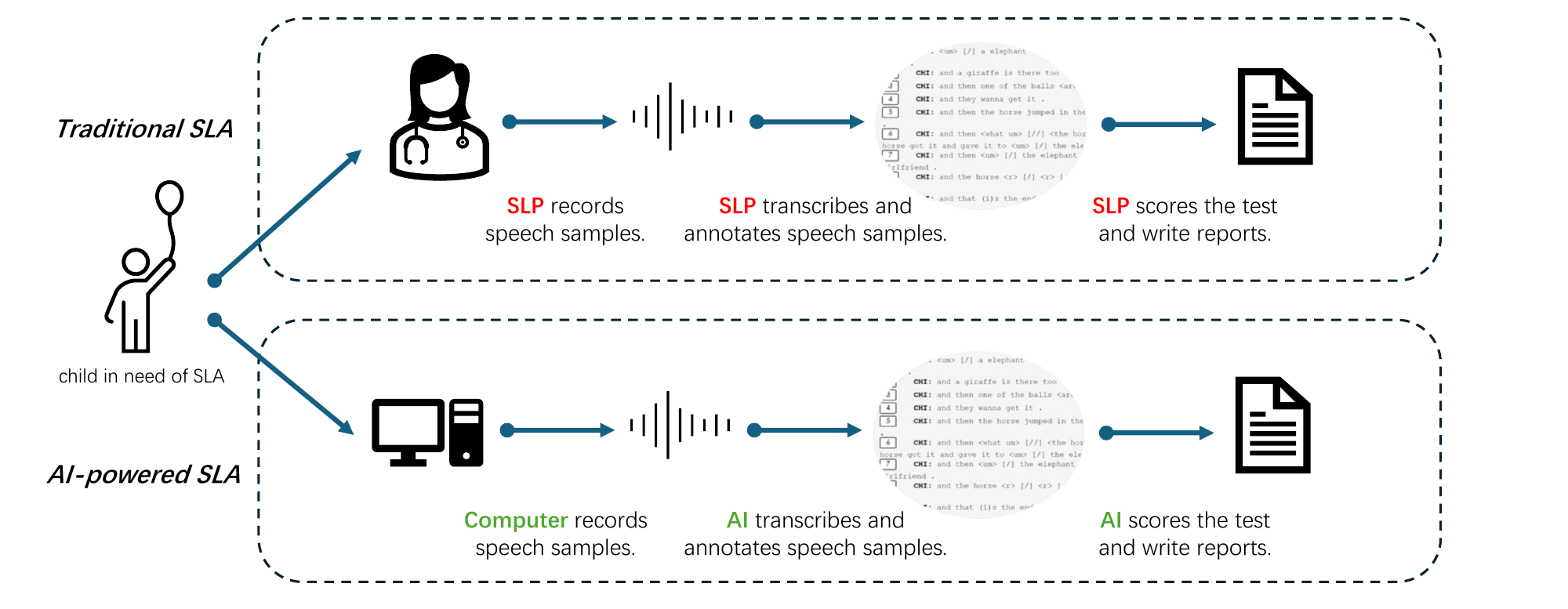}
  \caption{A generic pipeline for automated speech and language assessments in the future. Children in need of SLA will use a local computer to perform all required components of the test which ensures privacy and usability, and such remote SLA will also alleviate the burden of SLPs, helping them focus on the treatment of children in need.}
  \label{fig:pipeline_SLA}
\end{figure*}
\section{Introduction}

As arguably one of the most important abilities for human beings, speech plays a vital part in our lives. While speech seems to be used majorly for communication among adults,  its significance transcends communication for children, and 
children with speech disorders face unique challenges that can have enduring consequences if left unaddressed \cite{Bashir_Scavuzzo_1992}. 
Studies have shown that speech is not a mere means of expressing thoughts for children. Rather, it is the cornerstone of effective learning \cite{Young_Beitchman_Johnson_Douglas_Atkinson_Escobar_Wilson_2002,Tomblin_Zhang_Buckwalter_Catts_2000b}, appropriate social interaction \cite{Redmond_Rice_1998,Redmond_Rice_2002,Coster_Goorhuis-Brouwer_Nakken_Spelberg_1999} , and mental health \cite{Jerome_Fujiki_Brinton_James_2002,BEITCHMAN_WILSON_JOHNSON_ATKINSON_YOUNG_ADLAF_ESCOBAR_DOUGLAS_2001} during the formative years. 
Yet, it has been reported \cite{komodobrief} that about 9\% of children were diagnosed with speech disorders (SD) during assessments prior to the COVID-19 pandemic, and the number increased to 21\% in 2022. In total, roughly 1.2 million children in the United States were diagnosed with speech disorders in 2022 \cite{komodobrief}. The importance of speech 
and the huge number of affected children together call for immediate and effective screening and treatment.

Speech and language assessments (SLA) have long been essential tools that clinicians and practitioners use to facilitate early treatments for children with SD. SLAs were traditionally carried out by highly skilled speech-language pathologists (SLPs), but the burgeoning demand for efficient and scalable assessment methods has prompted a paradigm shift. Marked by the development of deep learning (DL), we are now 
able to integrate multiple AI-powered techniques with the traditional SLA pipeline, as shown in Figure \ref{fig:pipeline_SLA}, promising a revolutionary leap forward in the field of speech and language evaluation.

In this position paper, we bring forth three contributions to the SLA and DL community. First, we outline and survey existing techniques that could be used in this automated SLA pipeline, with a focus on automatic speech recognition (ASR). We will showcase the current progress in adapting ASR models to children's speech. Second, we discuss popular SLAs and their automated prototypes. We will show that the envisioned pipeline in Figure \ref{fig:pipeline_SLA} is actually not far from realization. Third, we highlight some of the practical concerns and solutions regarding the accessibility of the AI-powered pipeline and the privacy concerns on  user data.


\section{Automatic Speech Recognition}
In this section, we explore the fundamental technology that drives the automated SLA: automatic speech recognition (ASR). ASR takes a piece of audio and transcribes it into textual formats that humans visually understand. The output format is majorly in words, but other variations such as phonemes or characters also exist.

\subsection{Adult ASR}
The majority of works in the field of ASR focus on adult speech, in part due to the relatively broader impacts and easier acquisition of training data. Powered by massive datasets 
such as CommonVoice \cite{commonvoice}, 
ASR models such as Wav2Vec2 \cite{w2v2} and Conformer \cite{gulati2020conformerconvolutionaugmentedtransformerspeech} showcase excellent transcription quality for adults. More recently, the Whisper model \cite{radford2022whisper,bain2022whisperx} has revolutionized the ASR domain with large-scale weak-supervise training and has achieved high robustness towards any adult speech.

\begin{table}[b]
\begin{tabular}{l|l|l|l}
\hline
Model                        & Librispeech & ENNI & MyST \\ \hline
Wav2vec2-conformer           &    1.81 & 40.99 & 22.16\\
Wav2vec2                     &   2.64 & 43.89 & 25.65 \\
Whisper-small                &   3.30  & 16.34 & 17.57 \\
Nemo - STT                   & 1.67   & 32.41 & 16.64\\
 \hline
\end{tabular}
\caption{Performance (WER in percentage) of some recent ASR models on some representative speech datasets.}
\label{tab:asr_perf}
\end{table}

\subsection{Adapting to Children Speech}
However, even with the success of the Whisper model, obtaining high-quality ASR models that serve children remains a challenging problem. Children exhibit distinctively different characteristics compared to adults, including rapid changes in pitch, articulation patterns, and vocabulary development, and ASR models that work well with adults fail to excel with children \cite{Bhardwaj2022reviewASR}. Here, we provide some representative results of evaluating different ASR models on children's speech datasets in Table \ref{tab:asr_perf}. In particular, we choose the best-performing models -- Conformer \cite{gulati2020conformerconvolutionaugmentedtransformerspeech}, wav2vec2 \cite{w2v2}, Whisper \cite{radford2022whisper}, and Nvidia NeMo STT \cite{Harper_NeMo_a_toolkit} -- on the Librispeech dataset \cite{librispeech} and evaluate their performance on the two children datasets, ENNI \cite{liu2024fasaflexibleautomaticspeech} and My Science Tutor (MyST) \cite{pradhan2023science}. As we can see from these models' word error rate (WER), they suffer from significant accuracy degradation when applied to children's speech. Even the robust Whisper model shows a non-negligible difference that hinders its practical usage.

\begin{table}[b]
\newlength\q
\setlength\q{\dimexpr .12\textwidth -2\tabcolsep}
\noindent\begin{tabular}{p{0.2\textwidth}|p{0.1\textwidth}|p{0.1\textwidth}}

\hline
Model                        & ENNI & MyST \\ \hline
Whisper-small           &  16.34  & \textbf{17.57} \\
Whisper-small (FT)           & \textbf{7.04}   & 29.50\\ \hline
Whisper-medium                     & 17.19   & \textbf{16.67} \\
Whisper-medium  (FT)                   &  \textbf{4.88}  & 20.31 \\\hline
Wav2vec2 Large                      & 43.89   & \textbf{25.65} \\
Wav2vec2 Large (FT)                    &  \textbf{24.73}  
& 34.30
\\\hline
\end{tabular}
\caption{WER\% of some recent ASR models after fine-tuning (FT) on the ENNI and MyST speech dataset.}
\label{tab:asr_perf_ft}
\end{table}
To prevent accuracy degradation and ensure precise processing and analysis of children's speech for assessments, the adaptation of the 
 ASR models become a necessity. The prevalent adaptation technique employs various fine-tuning methods, with the most popular one being the low-rank approximation (LoRA) \cite{hu2021loralowrankadaptationlarge}. As shown in Table \ref{tab:asr_perf_ft}, we can see that the WERs of the ASR models decrease by a significant margin after fine-tuning. However, similar to the results obtained from \cite{attia2024kidwhisperbridgingperformancegap}, we see that the generalization ability (as measured by the performance fine-tuning on ENNI and testing on the MyST dataset) decreases after fine-tuning.

\subsection{Training Directly from Children Speech}

Recently, some researchers have taken a novel path different from fine-tuning. Similar to Whisper \cite{bain2022whisperx}, they hypothesize that pre-training directly from children's speech corpora will provide more benefits and bring more accurate and robust ASR models to the community. \cite{Li_2023} has shown promising results in improving speaker diarization and vocalization classification accuracy by pretraining the wav2vec2 model on massive children's home audio. A work-in-progress model that is directly trained from massive children's speech also preliminarily shows high accuracy and robustness in children's speech \cite{haolong}.

\section{Automatic Screening}
\subsection{Traditional Approach of SLA}
The conventional approach to SLA, characterized by manual assessments conducted by SLPs, has been a cornerstone in understanding linguistic capabilities. However, the inherent limitations of this approach include time-intensive procedures and resource demands, which delay the service for many children in need. In particular, as illustrated by the generic pipeline in Figure \ref{fig:pipeline_SLA}, after SLPs obtain speech samples from a child seeking SLA, they need to manually transcribe and annotate the speech samples. It is reported that for every one minute of speech sample, it takes seven to eight minutes for an experienced SLP to convert it to Systematic Analysis of Language Transcripts (SALT) format \cite{Miller_Andriacchi_Nockerts_2016}. Depending on the comprehensiveness of tests issued by SLP, the speech sample duration varies from 7-8 minutes for ENNI \cite{ENNI} to 30-45 minutes for core language assessments in CELF-5 \cite{CELF}. The extensive transcription and annotation time introduced in this stage hinders efficient testing for more children, but they could be potentially automated by deep learning models and AI-powered frameworks. Also, after proper annotations of the speech samples, SLPs will need to follow the manual and give out a score or a report on the child's speech ability. The artifact varies across different tests, but some of them are still time-consuming. For example, after the CELF-5 test \cite{CELF}, the SLP will issue a detailed report on the child's performance. While a template has already automated some parts of the report, it still requires some effort. 

\subsection{End-to-End System}
Currently, several end-to-end (E2E) evaluation systems have been proposed and implemented to evaluate children's speaking ability automatically. Various frameworks have been proposed to automatically classify children into typical development (TD) and speech-language impairment (SLI) groups, or similar variations. From the early statistical tests like \cite{gong16_interspeech} to the latest transformer-based models \cite{johnson23_interspeech,qin2024depression}, the claimed accuracy for identifying children with potential SLI problems has been raised to greater than 95\%. Similarly, there have been multiple consumer-level software from some companies that try to achieve similar goals, such as Ambiki \cite{Dias_2023b} and Smart Ears \cite{Smarty_Ears_2023}. While all aforementioned works and software show promising potential in automating the SLA and providing accessible remote services, they all either rely on black-box models or lack scientific explanations that hinder the interpretation of the results.

\subsection{Explanable System Derived from SLAs}
SLA for children is a sensitive area that requires careful attention and strong explanability. The cost of a false negative (that fails to screen a child with speech disorders) is much higher than typical deep-learning tasks. Because of that, frameworks that follow the guidance of SLPs and use tests developed by SLPs are favored over black-box models. As mentioned earlier, SLPs have come up with many forms of SLA tests, each looking at different aspects of children's speech, but all of the tests require extensive human labor, hindering their usage for most people.

We are working towards automating those tests and alleviating the workload of SLPs. Here, we report our progress and provide some preliminary results. In particular, we design and implement the automated Redmond Sentence Recall test (AutoRSR) \cite{Redmond_Ash_Christopulos_Pfaff_2019}. The AutoRSR follows the same process as the manual version, where it transcribes the children's speech sample with WhisperX \cite{bain2022whisperx}, segments the sample, aligns the transcription with the prompts using BertAlign \cite{bertalign}, applies a modified Levenshtein distance algorithm to count the errors and scores the children's speech sample. 

\begin{figure}[t]
  \centering
  \includegraphics[width=0.34\textwidth]{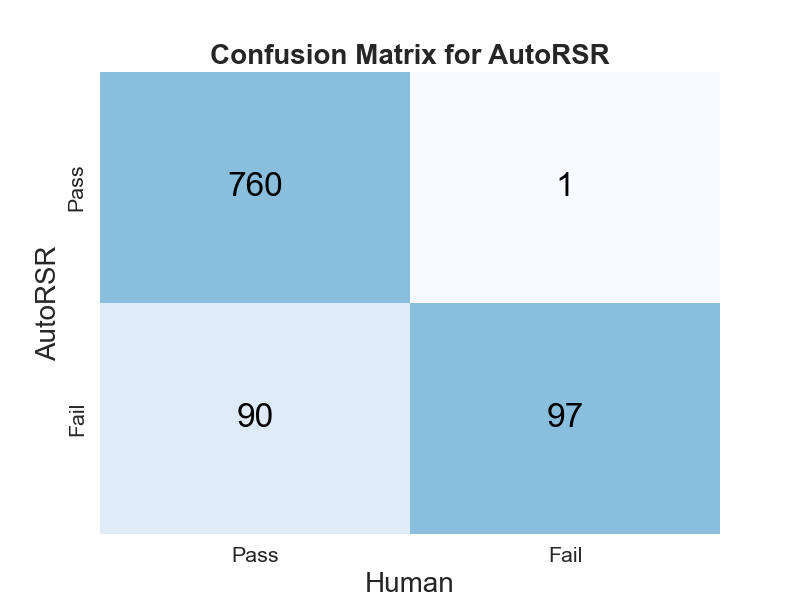}
  \caption{The confusion matrix on AutoRSR vs Human. Pass/Fail indicates whether the child passes/fails the RSR test. Cases when AutoRSR passes and Human fails the child are false negatives.}
  \label{fig:conf}
\end{figure}

\begin{figure}[!ht]
  \centering
  \includegraphics[width=0.36\textwidth]{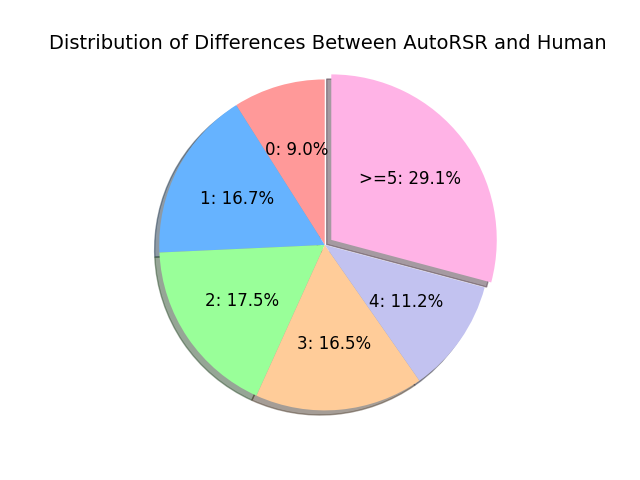}
  \caption{The distribution of differences between the automated RSR and human scorers. We can see that there is still a subtle difference between the scores even though the final classification is mostly correct.}
  \vspace{-3mm}
  \label{fig:pie_chart}
\end{figure}

We evaluate the performance of the AutoRSR software on the 948 samples with children's ages provided by the original authors of RSR, and the results are summarized in the confusion matrix in Figure \ref{fig:conf}. On the available samples, the AutoRSR achieved 90.4\% accuracy with only 1 false negative. Although the accuracy is high, we also observed that the Whisper model scores are harsher than those of SLPs. On average, Whisper scores are three points lower than human scores, and we provide a detailed distribution of score differences in Figure \ref{fig:pie_chart}. After consulting SLPs, we conclude this difference to be attributed to two factors: 1. Whisper sometimes makes mistakes and transcribes incorrectly; 2. SLPs transcribe with more context, causing them to guess some words when the child is pronouncing unclearly. Both factors would be alleviated by fine-tuning the model with RSR data (but as discussed above, at a potential cost of generalization).

However, despite the success of the automated RSR framework, it is noteworthy that generic automation of SLP-guided tests still faces numerous challenges. The biggest challenge among them is the need for an accurate phonetic ASR model that captures phoneme-level mistakes that children tend to make. While the RSR test does not require careful annotations on those phonetic mistakes, other tests such as the Edmonton Narrative Norms Instrument (ENNI) test require SALT annotations and emphasize the correct language regularization \cite{ENNI}. As a result, current state-of-the-art Whisper-based models are no longer suitable for automating the ENNI test, because Whisper often infers words, leading to an auto-correct regularization that deviates from the intended purpose. 

\subsection{Report Generation}
The final step in some of the tests such as CELF \cite{CELF} requires the SLP to issue a report. As mentioned earlier, such a process is very time-consuming with little expertise required. Given the development of LLMs, we believe that they could help make this process fully automated. While there have not been works investigating the potential of LLMs generating SLP reports, similar works have been successful in asking LLMs to generate financial reports \cite{zhao2024revolutionizingfinancellmsoverview}, extracted clinical notes \cite{li2024automatedclinicaldataextraction}, and SOAP notes \cite{hu2021loralowrankadaptationlarge}. We believe that SLP reports could also be achieved in similar ways with an adequate knowledge base and appropriate usage of RAG.
\section{Practical Concerns on Availability and Privacy of Automated SLA}
In this section, we discuss some of the practical concerns that limit the usage of automated SLAs under resource-constrained settings. Those concerns and their solutions are particularly important for the deployment of AI-powered SLAs in less developed regions.

\subsection{Availability of DL Models due to Privacy Concerns}
Children's speech is highly sensitive data that requires a high level of privacy. Due to the uncontrollable nature of cloud services \cite{qin2024empiricalguidelinesdeployingllms} and the unacceptable cost of secure computation \cite{crypten2020,xu2024outofdistributiondetectiondeepmulticomprehension}, the most favorable method for the automated SLA is to deploy ASR models and other necessary models onto edge devices. However, edge ASR presents a unique set of challenges that need to be addressed, including the RAM limitations, the tradeoff between model quantization and performance, and computation overhead and delays, all of which will affect the practical performance of the automated SLA. 

Numerous works have tried to address different individual challenges. For example, PI-Whisper provides fine-grained fine-tuning and inference via individualized profiling according to additional characteristics of the speakers and archives high accuracy using smaller models \cite{nassereldine2024piwhisperadaptiveincrementalasr}. Whisper-KDQ provides a working implementation of a distilled and quantized Whisper model that fits into edge devices while even increasing the accuracy \cite{shao2023whisperkdqlightweightwhisperguided}. Transformer accelerations that reduce inference time, such as flash attention \cite{dao2022flashattentionfastmemoryefficientexact}, ONNX runtime with process-in-memory \cite{ONNX_PIM}, and even optimized edge transformer architectures \cite{bergen2021systematic,quadranet}, could also help the automated SLA on the edge. However, these works do not integrate with each other, and a unified fast, accurate, and low-resource ASR model that serves the need is yet to be explored.

\subsection{Fairness towards Marginalized Group}

Moreover, the fairness of ASR models and the SLA framework needs to be researched, especially when facing marginalized groups. While the bias of ASR models has been researched for adults, in particular, African American accents \cite{aae_bias, Mengesha_Heldreth_Lahav_Sublewski_Tuennerman_2021, alwan_2024_aae}, there has been little research involving the speech of children. At the same time, how the fine-tuning process affects each group of speakers is seldomly studied, with only PI-Whisper showing some results, indicating that profile-based fine-tuning helps the fairness of ASR \cite{nassereldine2024piwhisperadaptiveincrementalasr}.
\section{Conclusion}

In this paper, we have explored the intricacies of ASR techniques tailored for children, discussed the development and implementation of an automated speech-language assessment pipeline, and addressed the critical privacy concerns associated with edge ASR technology. Our high-level discussions and practical implementations highlight the promising advancements in these areas and their potential to benefit children significantly.

We have demonstrated that ASR systems can be effectively adapted to account for the unique speech patterns and linguistic behaviors of children \textbf{on word-level}, but the \textbf{generalizability} of children ASR models and \textbf{phonetic} ASR models remain an open challenge in the area. Similarly, with the current ASR models, we can prototype some (simple) automated speech-language assessments, but more complex SLAs require stronger ASR models to achieve desirable practicality. Furthermore, the deployment of private ASR models on the edge deserves more research focus, particularly on the direction of integrating multiple optimization methods and fine-tuning more fair and accessible models.

\section{Acknowledgements}
This material is based upon work supported under the AI Research Institutes program by National Science Foundation and the Institute of Education Sciences, U.S. Department of Education through Award \# 2229873 - National AI Institute for Exceptional Education. Any opinions, findings and conclusions or recommendations expressed in this material are those of the author(s) and do not necessarily reflect the views of the National Science Foundation, the Institute of Education Sciences, or the U.S. Department of Education. The work is also supported in part by the NSF Grants \# 2235364 and \# 2329704.

\bigskip

\bibliography{aaai24}

\end{document}